\documentclass[12pt]{article}
\usepackage{latexsym}
\newcommand{\bba}{\begin{eqnarray}}
\newcommand{\eea}{\end{eqnarray}}
\newcommand{\bb}{\begin{equation}}
\newcommand{\ee}{\end{equation}}
\newcommand{\bban}{\begin{eqnarray*}}
\newcommand{\eean}{\end{eqnarray*}}

\newcommand{\dddotV}{\buildrel{...}\over V}
\newcommand{\ddddotV}{\buildrel{....}\over V}
\begin{document}

\begin{titlepage}

\title{Vacuum polarization in the
Szekeres Class of colliding plane wave space-times}

\author{\null}
\author{  {\sc Miquel  Dorca}\footnote{
E-mail: mdorca@rainbow.uchicago.edu}$\;\;$\footnote{
Present address: Box 1843, Department of Physics, 
               Brown University, Providence RI 02912
                                              } \\ 
          {\small\em  Enrico Fermi Institute}   \\
          {\small\em  The University of Chicago}  \\
          {\small\em  5640 Ellis Avenue}        \\
          {\small\em  Chicago IL 60637}    }

\date{February  19, 1998}

\maketitle

\begin{abstract}

We study the quantization of a scalar field on a classical background
given by the {\em Szekeres Class of
solutions}, which represent the collision of
two gravitational plane waves with constant polarization. These
solutions consist of two approaching gravitational plane waves moving
in a flat background and an interaction region which always contains a
curvature singularity. 
Following a suitable approximate procedure, introduced in a previous
paper, we propose a way to compute the vacuum 
expectation value of the stress-energy
tensor throughout the causal past region of the collision center in the
quantum state which corresponds to the vacuum before the arrival of
the waves.

\end{abstract}



\vskip 0.5 truecm
\noindent
EFI-98-08

\end{titlepage}

\section{introduction}

Gravitational waves in General Relativity are considered as perturbations
of the space-time geometry which propagate with the speed of light.
Contrarily to the behaviour of electromagnetic waves in flat space,
when gravitational waves pass through each other there is a non-linear
interaction. Of course, such a difference arise from the fact that,
unlike Maxwell's equations, Einstein's equations are highly non-linear.
The situation
is very different in the linearized theory of gravity where gravitational
waves superpose without interaction. Even in the full Einstein-Maxwell
theory, Maxwell's equations remain linear indicating non direct
electromagnetic interaction between two pure 
electromagnetic  waves. In that case, however, 
there is a non-linear interaction of the waves through the gravitational 
field generated by their electromagnetic energy. Although very small,
the magnitude of such
an interaction is similar to the magnitude of the interaction between
pure gravitational waves.

In order to find exact solutions of Einstein's equations describing plane
wave collisions
it is necessary to introduce several simplifying
assumptions. Perhaps, the most restrictive simplification would be the
assumption that the interacting waves have plane symmetry. Indeed,
exact {\em gravitational plane waves} are very simple
time dependent plane symmetric solutions of Einstein's
equations \cite{bel26}. That may not seem a very restrictive assumption
if we intuitively think of plane fronted
waves as approximations to spherical waves at large distances from
their sources. However, the global features of the two type of waves
are very different. In fact, exact gravitational plane waves exhibit
two main surprising global features, namely: i) the absence of a
global Cauchy surface, which is a consequence of the focusing effect 
that the waves exert on null rays  \cite{pen65}, ii) the presence of
a coordinate singularity on some hypersurface
behind the wave. Such a singularity may be physically understood in 
terms of {\em caustics} that are formed in the regions where the null
rays are focused \cite{pir89}. For pure gravitational plane waves, a whole
three-dimensional congruence of parallel null geodesics may be focused
into  a line. For pure electromagnetic plane waves, the region of focusing
may be a simple event. From this property it follows that we can 
conveniently consider the inverse of the focusing time as a measure of
the strength of the wave. In particular, for a pure electromagnetic wave
such inverse time equals the electromagnetic energy per unit surface of the
wave. 

Beyond the assumption that the wave front is plane, the
imposition of plane symmetry also requires that the magnitude of the 
wave is constant all over the entire plane. Furthermore, if we only consider
{\em head on} collisions, then it makes sense to impose the condition
of plane symmetry globally. Actually, it may seem that there is not loss
of generality if we concentrate only on head on collisions. The reasons
are that the particular case of
plane waves propagating in the same direction is trivial since the waves
do not interact, they simply superpose linearly \cite{bon69,aich71}.
Also, we may think that it is always possible to make a Lorentz 
transformation in order to include oblique collisions. However, the
assumption of global plane symmetry would severely blur the physical
interpretation of the solutions obtained in this particular way.

Nevertheless, besides their simplicity, 
one expects that exact plane waves may be relevant for 
the study of the strong time
dependent gravitational fields that may be produced in the collision of black 
holes \cite{dea79,fer80} or to
represent travelling waves on strongly gravitating cosmic strings 
\cite{gar89-90}. In recent years these
waves have been used in classical general relativity to test some conjectures 
on the stability of Cauchy
horizons
\cite{ori92,yur93}, and in string theory to test classical and quantum string 
behaviour in strong
gravitational fields \cite{veg84-90,veg91,jof94}. Their interest also stems 
from the fact that plane waves are a
subclass of exact classical solutions to string theory 
\cite{ama84-88,hor90,tse93fes94rus95}. 

It is known that when waves are coupled to quantum fields there is
neither vacuum polarization nor the spontaneous creation of particles.
In that sense they behave 
very much as electromagnetic or
Yang-Mills plane waves in flat space-time \cite{des75,gib75}. 
Still the 
classical focusing of geodesics has a
quantum counterpart: when quantum particles are present the quantum field 
stress-energy tensor between
scattering states is unbounded at the Cauchy horizon, i.e. where classical 
test particles focus after colliding
with the plane wave
\cite{gar91}. This suggests that the Cauchy horizon of plane waves may be 
unstable under the presence of quantum
particles. The classical instability of the null Cauchy horizons of plane 
waves is manifest when non-linear
plane symmetric gravitational radiation collides with the background wave, 
i.e. when {\it two plane
waves collide}. In this case the focusing effect of each wave distorts the 
causal structure of the space-time
near the previous null horizons and either a spacelike curvature singularity 
or a new regular Killing-Cauchy
horizon is formed. However, it is generally believed that the Killing-Cauchy 
horizons of the colliding plane
wave space-times are unstable in the sense that ``generic" perturbations will 
transform them into spacelike
curvature singularities. In fact, this has been proved under general plane 
symmetric perturbations
\cite{yur88-89}. Also exact colliding plane wave solutions with classical 
fields are known that have
spacelike curvature singularities and which reduce, in the vacuum limit, to 
colliding plane wave solutions
with a regular Killing-Cauchy horizon
\cite{cha87}.

Note that the singularities 
derived from plane wave collisions are not the 
result of the collapse of matter but the result of the non-linear
effects of pure gravity. Note also that
the presence of a Killing-Cauchy horizon in a colliding
plane wave space-time implies a breakdown of predictability since the
geometry beyond the horizon is not uniquely determined by the initial
data posed by the incoming colliding waves. 
In fact, these
extensions exhibit a wide variety, they may or may not contain
space-like curvature singularities. Even they may contain time-like
singularities, which can be avoided for certain observers in a
space-like motion (see \cite{gri91} for a review on the subject).

The interaction of quantum fields with colliding plane waves was first
considered by Yurtsever \cite{yur89} for the singular Khan-Penrose
solution \cite{kha71}, which describes the collision of two plane impulsive
gravitational waves. In that case,
an unambiguous ``out'' vacuum state was possible to define in a relatively
simple way. More recently, Dorca and Verdaguer \cite{dor93,dor94} 
noticed that the presence
of a Killing-Cauchy horizon in a non singular colliding plane wave 
space-time could be used to define an unambiguous ``out'' vacuum state
related to the preferred Hadamard state introduced by Kay and Wald in more
generic space-times with Killing-Cauchy horizons \cite{kay91}. 
With this premise,
Dorca and Verdaguer studied the interaction of quantum fields in a 
particular non-singular colliding plane wave space-time, the
interaction region of which was isometric to a region 
inside the event horizon of
a Schwarzschild black hole \cite{fer87cha86,hay89}. Later on the same
premise was applied by Feinstein and Sebasti\'an \cite{fei95} 
to the Bel-Szekeres
solution \cite{bel74}, which represents the head 
on collision of two electromagnetic
plane waves with an interaction region isometric to the Bertotti-Robinson
universe \cite{rob54ber59} filled with an uniform electric field.
In all these examples it was found that the initial state, defined
to be the vacuum state in the flat region before the arrival of the waves,
contained a spectrum of ``out'' particles consistent, in the
long wavelength limit,  with a thermal spectrum with a temperature inversely
proportional to the focusing time of the waves.

A further step in the study of the interaction of quantum fields with
colliding plane waves is the computation of the expectation value of
the stress-energy tensor. However,
the non perturbative evaluation of the stress-energy 
tensor of a quantum field in a
dynamically evolving space-time is generally a difficult task.
Even when the exact modes of the quantum field equation are known it may not 
be possible to perform the mode
sums in order to get the quantum field two point function or, more precisely, 
the Hadamard function, which is
the key ingredient in the evaluation of the stress-energy tensor.
This problem was first considered by
Yurtsever \cite{yur89} for the Khan-Penrose solution \cite{kha71}.
In that case it was possible to determine the behaviour of the stress-energy
tensor near the singularity of the interaction region.
It was shown that for the conformal coupling case (i.e. $\xi =1/6$) the
energy density and two of the principal pressures  were 
positive and unbounded towards the singularity. 
This problem has been also considered by Dorca and Verdaguer in 
the mentioned above non-singular colliding plane wave spacetime with an
interaction region isometric to an interior region of a Schwarzschild
black hole. As in the case of Yurtsever for
the Khan-Penrose solution, the expectation value of the stress-energy
tensor was calculated in the state representing the Minkowski vacuum
in the flat region before the arrival of the waves. This value
was first computed in a region close to both the Killing-Cauchy
horizon and the topological singularities, the {\em folding singularities}, 
that the colliding plane wave space-time contains \cite{gri91}.
In that particular region, the calculations were simplified due to
the blueshift effect on the energy of the initial quantum modes as
they reached the Killing-Cauchy horizon \cite{dor96}. In a recent work
\cite{dor97}, an approximation procedure was proposed by the author
in order to calculate such an expectation value  throughout the causal
past region of the collision center. That approximation has been also
applied by the author in the Bel-Szekeres space-time \cite{dor98a}.
In all of these calculations, it was found that
the stress-energy diverged as the Killing-Cauchy horizon was approached.
The rest energy density was positive and unbounded towards the horizon.
Two of the principal pressures were negative and of the same order of 
magnitude of the energy density. It was also pointed out that such a
behaviour suggested that the non singular Killing-Cauchy horizon is
indeed unstable under quantum perturbations and a curvature singularity
would be the general outcome of a generic plane wave space-time
when backreaction is taking into account.
Note that this is a non perturbative
effect, it is the result of the nonlinearity of gravity, since gravitational 
waves in the linear approximation
do not polarize the vacuum. In fact the vacuum stress-energy tensor of a 
quantum field in a weakly inhomogeneous
background was computed by Horowitz \cite{hor80}, and it is easy to see that 
such tensor can be written in
terms of the linearized Einstein tensor only \cite{cam94}, which vanishes for 
gravitational waves \cite{can80sci81,fro85tho86}.

The plan of the paper is the following. In section 2 the geometry of plane
fronted waves and colliding plane waves  is
briefly reviewed. In section 3 the mode solutions of the scalar 
field equation are given throughout the causal past of the
collision center in the four different 
regions of the space-time. It is explicitly seen that 
an exact expression for these mode
solutions can be easily found everywhere but in the interaction region.
However, following the directions of two previous works \cite{dor97,dor98a},
we will give an adequate approximation for the field solutions in the
interaction region. Such an approximation will allow us to recover
the singularity pattern of a Hadamard function, which is
essential in order to perform a correct regularization.
In section 4 a brief summary of the main {\em point-splitting} regularization
formulae is given. Finally, in section 5, a procedure to calculate the 
Hadamard function is introduced. 
Also, are given the basic details to compute the
vacuum expectation value of the stress-energy tensor in the causal
region of the collision center. A summary 
and some consequences of the results are discussed in section 6. Also,
some bitensor covariant expansions  have been stored in the Appendices.

\section{Description of the geometry}

In this section we  start with a review of the geometrical
properties of gravitational plane waves, and we follow with a
geometrical analysis of the head on collision of two gravitational
plane  waves.

\subsection{Gravitational and electromagnetic plane waves}

The appropriate notation to describe plane waves is that of Newman and
Penrose (1962) \cite{new62}. We introduce a thetrad on null vectors,
namely, two real vectors $n^\mu$, $l^\mu$, and a complex null vector
$m^\mu$ and its complex conjugate ${\bar  m}^\mu$, such that
$n^\mu l_\mu =1$, $m^\mu{\bar m}_\mu =-1$, and they satisfy the
completeness relation,

\[g_{\mu\nu}=l_\mu n_\nu +n_\mu l_\nu -m_\mu{\bar m}_\nu -{\bar m}_\mu
m_\nu .\]
Then we can write the ten independent components of the Ricci and the
ten independent components of the Weyl tensor in this thetrad basis.
This turns to be physically convenient since the components of the
Weyl tensor in the thetrad basis have a well established meaning of
Coulomb components or
transverse and longitudinal wave components in the null directions
$n^\mu$, $l^\mu$ \cite{sze65}. This
interpretation is also very useful to analyze the interaction of two
of such waves since it is possible to align the basis vectors
$n^\mu$ and $l^\mu$ with the propagation directions of the two waves.
Furthermore,
if we consider the interaction between transverse waves, the
problem reduces to find the interaction between the respective
transverse Weyl components.

In the context of plane waves we introduce the class of {\em
pp-waves}, i.e plane fronted gravitational waves with parallel
rays. They are defined by the property that they admit a covariantly
constant null vector field, which is possible to interpret as the
rays of the wave and it may be identified with the thetrad vector
$l^\mu$. There exist a family of 2-surfaces orthogonal to $l^\mu$ that
may be interpreted as wave surfaces \cite{kun61}.
It is convenient to use a null coordinate $u$, defined by $l_\mu
=u_{,\mu}$, and the metric can be written in the {\it Kerr-Shild form}
as,

\bb  ds^2=2\, dudr+H(u,X,Y)\, du^2-dX^2-dY^2, \label{eq:ppwave}\ee
where the coordinates $X$ and $Y$ span the wave surfaces. These type of metrics
are either of algebraic type N, or conformally flat and the only non-zero 
component of the Weyl tensor, in the thetrad basis, is
the transverse wave component in the $l^\mu$
direction. The modulus term of this component can be conveniently
interpreted as the {\em amplitude} of the wave and its phase term as
the {\em polarization} of the wave.

If we suppose that these space-times are solutions of Einstein's
equations, the function $H$ in the metric (\ref{eq:ppwave}) must
satisfy the differential equation,

\[ H_{,XX}+H_{,YY}=0.  \]
Recall that this is a linear differential equation which means that
distinct solutions of pp-waves (\ref{eq:ppwave}) may be superposed,
i.e. two independent pp-waves, propagating in the same direction do
not interact \cite{bon69,aich71}.
Another remarkable property of pp-waves, which was pointed out by
Yurtsever \cite{yur88}, states that any gravitational wave space-time
that is flat before the arrival of the wave and returns to a perfect
flatness after the wave passes, i.e. a {\em sandwich wave}, it is
necessarily a pp-wave.

In {\em Einstein-Maxwell theory}, the particular class of
plane-symmetric waves
are defined to be pp-waves in which the curvature field components are the same
at every point of the wave surfaces, i.e. the
transversal $X, Y$-planes. There is only one non-null component
of the Ricci tensor and, as mentioned above, only one non-null component
of the Weyl tensor. Both components depend only on second
derivatives of the function $H(u,X,Y)$ with respect to the transversal
coordinates $X$ and $Y$. Therefore, the plane symmetry condition implies
that $H(u,X,Y)$ is quadratic in $X$ and $Y$, i.e.,

\[ H(u,X,Y)=h_{11}(u)X^2+2\, h_{12}(u)XY+h_{22}(u)Y^2. \]
In that case the non null components of the Ricci tensor, in the
thetrad basis, is given by
the combination $(h_{11}+h_{22})/2$ and the non null component
of the Weyl tensor, in the
thetrad basis, i.e. the transversal wave component in the direction
$l^\mu$, is given by the combination,
${1\over 2}\left( h_{11}-h_{22}+i\, 2\, h_{12}\right)$.
Then, depending on the values of $h_{ij}(u)$, we will distinguish
between two type of waves, namely, i) {\em pure gravitational waves}, when
$h_{11}=-h_{22}$, and therefore the Ricci tensor is exactly zero and
we have a vacuum solution of Einstein's equations. If in addition
$h_{11}$ is proportional to $h_{12}$, then the wave has constant
linear polarization. ii) {\em pure electromagnetic waves} when
$h_{11}=h_{22}$ and $h_{12}=0$, and therefore the Weyl tensor is
exactly zero.

In order to consider the collision and interaction of plane waves it
is first convenient to change to {\em Rosen-like}
coordinates ($u$, $v$, $x$, $y$), where $u$ and $v$ are two null
coordinates, $x$ and $y$ two transverse coordinates, and the line
element (\ref{eq:ppwave}) can be transformed into,

\bb ds^2=2\, dudv-{\rm e}^{-U}\left(
{\rm e}^V\cosh W dx^2-2\sinh W dx dy+{\rm e}^{-V}\cosh W dy^2
\right), \label{eq:pwrosen}\ee
where $U$, $V$ and $W$ are functions of $u$ only. In the case of
linear polarization, it is always possible to set $W=0$.

However, the line element (\ref{eq:pwrosen}) in Rosen form, always
contains a coordinate singularity on some hypersurface behind the
wave. This singularity appear as a consequence of the focusing
effect which plane waves exert on null geodesics. Physically it may be 
interpreted as a {\em caustic} produced in the region of
null ray  focusing \cite{pir89}. Related to this focusing property of plane
waves, as was first pointed out by Penrose \cite{pen65}, 
there is the fact that plane wave
space-times do not contain global Cauchy surfaces, i.e it is not
possible to set up initial values for a plane wave on any global
space-like surface which lies entirely at the causal past of the wave
front.

\subsection{Colliding waves}

The colliding plane wave space-time consists of two
approaching waves, regions II and III, in a flat background, region
IV, and an interaction region, region I. The two waves move in the
direction of two null coordinates $u$ and $v$, and since they have
plane symmetry in the direction of the transversal coordinates $x$ and
$y$, the interaction region retains a
two-parameter symmetry group of motions generated by the Killing
vectors $\partial _x$ and $\partial _y$. At each point of the
interaction region, the null directions, along coordinates $u$ and
$v$, are orthogonal to the planes spanned by $\partial _x$ and
$\partial _y$. We may align the thetrad vectors $n^\mu$ and $l^\mu$
with these two null directions in such a way that,
$l_\mu =u_{,\mu}/L$ and $n_\mu =v_{,\mu}/N$, being $L$ and $N$ two
functions which do not depend on coordinates $x$ and $y$.

The general line element for the interaction region can be written in
{\em Szekeres form} as

\bb ds^2=2{\rm e}^{-M}dudv -{\rm e}^{-U}\left(
{\rm e}^V \cosh W +2\sinh W dx dy +{\rm e}^{-V}\cosh W dy^2
\right), \label{eq:dsI}\ee
where $M$, $U$, $V$ and $W$ are functions of $u$ and $v$ only. These functions
are constrained by Einstein's equations together with a suitable set
of initial conditions at the boundaries with the plane wave regions II
and III.
It is possible to set in general,

\bb {\rm e}^{-U}=a(u)+b(v), \label{eq:UI}\ee
with $a(u)$ and $b(v)$, two arbitrary {\em monotonically decreasing}
functions that may be considered piecewise $C^1$. Recall that the line
element (\ref{eq:dsI}) is rather convenient because it is similar to
that for a single plane wave in Rosen form (\ref{eq:pwrosen}), except
that the functions $M$, $U$, $V$ and $W$ are functions of both 
coordinates $u$ and $v$ in region I, of coordinate $u$ alone in region
II and of coordinate $v$ alone in region III. In
region IV all these functions are constant and may be set to be
zero by an appropriate coordinate parametrization.

The appropriate junction conditions across the null hypersurfaces
$u=0$ and $v=0$, are those of O'Brian and Synge \cite{obr52}. These
conditions demand that only the components,
$g_{\mu\nu}$, $g^{ij}g_{ij,0}$, $g^{i0}g_{ij,0}$,
should be continuous through the null hypersurfaces. Notice that
O'Brian and Synge conditions are weaker than the usual Lichnerowicz
conditions, which require that the metric tensor is $C^1$ and
piecewise at least $C^2$. In particular, Lichnerowicz conditions
exclude the possibility of {\em impulsive gravitational waves}.

The function $U$ is appropriately set by the boundary conditions
(see for instance \cite{gri91}):

\[ a(u)={1\over 2};\;\; u\leq 0,\;\;{\rm and}\;\; {\dot a}(0)=0, \]
\[ b(v)={1\over 2};\;\; v\leq 0,\;\;{\rm and}\;\; {\dot b}(0)=0. \]
Thus, the values of $\exp (-U)$ are $1$ in region IV, $1/2+b(v)$ in region
III, $1/2+a(u)$ in region II and $a(u)+b(v)$ in region I. The
statement that $U$ is an smooth function across the null boundaries
requires that the function $a(u)$ must have the same form in both
regions I and II. Similarly, the function $b(v)$ must have the same form in
regions I and III. Furthermore, being $a(u)$ and $b(v)$ monotonically
decreasing functions, it is always possible to express them in the
form,

\bb a(u)={1\over 2}-\Theta (u)\,\left( c_1\, u  \right)^{n_1},\;\;\;
    b(v)={1\over 2}-\Theta (v)\,\left( c_2\, v  \right)^{n_2}, 
\label{eq:abc1c2}\ee
for a certain positive values of the parameters $c_1$, $c_2$, $n_1$,
$n_2$ and where $\Theta (\zeta )$ is the usual step function.
Recall that this expressions for $a(u)$ and $b(v)$ apply globally. 
Observe that, since functions $a(u)$ and $b(v)$ are monotonically
decreasing, it is inevitable that a singularity, either a curvature
singularity or a Killing Cauchy horizon, at $a(u)+b(v)=0$ is
formed. The two constants $c_1$ and $c_2$ in (\ref{eq:abc1c2}) are a
measure of the strength of the waves. In fact, the larger these
parameters are, the sooner the singularity is produced.
It is also possible to use further rescaling on null coordinates $u$
and $v$ to set $c_1=c_2=1$, and with this rescaling, as depicted in Fig. 1,
we have a symmetric wave collision. Once a rescaling is used, to
set $c_1=c_2=1$, then a measure of the 
strength of the waves is absorbed into the
function $\exp (-M)$ in line element (\ref{eq:dsI}). In fact,
$\exp (-M)$ will be proportional to $L_1L_2$, where the two new 
parameters $L_1$ and $L_2$ are directly related to $c_1$ and $c_2$ as
$L_1=c_1^{-1}$ and $L_2=c_2^{-1}$.

Once a value for function $U$ is set, Einstein's equations together
with suitable initial conditions at the boundaries with the null
hypersurfaces $u=0$ and $v=0$, will determine the values for the
remaining functions, i.e $V$, $M$ and $W$. For the particular case of
linearly polarized plane waves, we may set $W=0$. In that case,  one possible
family of solution of Einstein's  equations which  
satisfy the boundary conditions is the {\em Szekeres Class of
solutions} \cite{sze72}, for which,

\bb{\rm e}^{ V(u,v)}= (a+b)^{p_1+p_2}
\left(\sqrt{1/2+b}+\sqrt{1/2-a}\right)^{-2p_1}
\left(\sqrt{1/2+a}+\sqrt{1/2-b}\right)^{-2p_2} ,
\label{eq:Vuv}\ee

\bb {\rm e}^{-M(u,v)}={(a+b)^{[(p_1+p_2)^2-1]/2}\over (1/2+a)^{p_2^2/2}
(1/2+b)^{p_1^2/2}}\left(
\sqrt{1/2-a}\sqrt{1/2-b}+\sqrt{1/2+a}\sqrt{1/2+b}
\right)^{-2p_1p_2}, \label{eq:Muv}\ee
where $p_1$ and $p_2$ are two real parameters related to the
parameters $n_1$ and $n_2$ in the definition of functions $a(u)$ and $b(v)$
(\ref{eq:abc1c2}) by,

\bb p_1^2=2\left(1-{1\over n_1}\right),\;\;\; 
   p_2^2=2\left(1-{1\over n_2}\right).
\label{eq:p1p2}\ee
In order to satisfy the boundary conditions, it is necessary that, 
$n_i\geq 2$ with $i=1,2$, and therefore the constants $p_i$ are restricted
to the range of values $1\leq p_i^2<2$.
For instance, the particular
values $p_1=p_2=1$ correspond to the Khan-Penrose solution \cite{kha71}.
In this family of solutions, there is always a scalar polynomial curvature
singularity in the interaction region on the surface $a(u)+b(v)=0$.

We will restrict in the following sections to the Szekeres Class of solutions.
The colliding plane wave space-times can be briefly described as
follows (see Fig. 1):
The general space-time contains four regions, namely: 
two single plane wave regions (regions II and III) moving in a Minkowski
flat background (region IV) and an interaction region (region I).
These four space-time regions
are separated by the two null wave fronts $u=0$ and $v=0$. Namely, the
boundary between regions I and II is  $\{0\leq u<1,\; v=0\}$, the
boundary between regions I and III is $\{u=0,\; 0\leq v <1\}$, and
the boundary between regions II and III with region IV is
$\{u\leq 0,\;v=0\}\cup\{u=0,\;v\leq 0\}$.
The interaction region is bounded by a scalar polynomial curvature
singularity at the surface $a(u)+b(v)=0$.
Region I meets region IV only at the surface $u=v=0$, 
and the plane wave region II or III meets the  singularity of region I 
at ${\cal P}=\{u=1,\; v=0\}$ or ${\cal P}'=\{u=0,\; v=1\}$, respectively. 
Observe that,
the single plane wave regions II and III always contain a singularity
at the hypersurface $u=1$ for region II and $v=1$ for region III.
For the particular case of the Khan and Penrose solution, which
corresponds to the values $n_1=n_2=2$, these singularities
are just coordinate singularities. For all the other cases, 
i.e. $n_1,n_2>2$, these singularities are  non-scalar curvature singularities
\cite{kon89}. However, in what follows we will refer generically to them as 
{\em folding singularities}. The reason for this terminology
comes from the behaviour of null geodesics near them.
Namely, only a null measure set of null geodesics
which enter into the plane wave regions II or III  hit the singularities,
all the rest are {\em folded} to end up into the interaction region.
Such a behaviour suggest that the whole singularity at $u=1$ in
region II (or $v=1$ in region III) should be identified with the surface
${\cal P}=\{u=1,\; v=0\}$ (or ${\cal P}'=\{u=0,\; v=1\}$)
in region I (see for instance \cite{dor93} for a 3-dimensional plot
of a space-time of this type).

\section{Mode propagation}

For simplicity we will consider in this section a massless scalar field,
which satisfies the usual Klein-Gordon equation,

\bb \Box\phi =0.\label{eq:kG}\ee
Following the directions of the approximation procedure introduced
in the previous works \cite{dor97,dor98a}, 
we will be interested in the value of the quantum field
$\phi$ all over the causal past region of the collision center. 
The reason is essentially because the calculations can be greatly
simplified in this region. We will
start with the field solution in the flat region prior to the arrival
of the waves, which is chosen to be the usual vacuum in Minkowski
space-time. This vacuum solution will set a well posed initial value
problem on the null boundary $\Sigma =\{u=0,\; v\leq
0\}\cup\{u\leq 0,\; v=0\}$, by means of 
which a unique solution for the field equation can be found throughout
the space-time, i.e., in the plane wave regions (regions II and III), 
and in the interaction region (region I).
However, although it is rather easy to find the solution of the field
equation in regions II and III
which matches smoothly with the boundary conditions, it turns out to be a
difficult problem for the interaction region. The reason is
essentially due to the intrinsic differences between the geometry of
the plane wave regions and the interaction region. In fact,
as mentioned in section 2, the plane
wave regions are either conformally flat or type N in the Petrov
classification, but the interaction region can be more generic.
We will refer, from now on, to this problem as {\em the mode
propagation problem}. 

We will consider the line element,

\bb ds^2=2{\rm e}^{-M(u,v)}dudv-{\rm e}^{-U(u,v)}\left(
{\rm e}^{V(u,v)}dx^2+{\rm e}^{-V(u,v)}dy^2
\right), \label{eq:dsG}  \ee
which applies globally to the four space-time regions, and where the
functions $U$, $V$ and $M$, can be directly read off
(\ref{eq:UI})-(\ref{eq:Muv}). Then, the
field equation can
be separated in a plane-wave form solution for the transversal
coordinates $x$ and $y$, with $k_x$ and $k_y$, respectively, as
separation constants. This plane-wave separation is just a trivial 
consequence of the translational symmetry of the space-time on the planes
spanned by the Killing vectors $\partial _x$ and $\partial _y$. The
field solution is thus,

\bb \phi (u,v,x,y)={\rm e}^{U(u,v)/2}\, f(u,v)\, {\rm e}^{ik_xx+ik_yy},
\label{eq:phiG}\ee
where the function $f(u,v)$ satisfies the following second order
differential equation,

\bb f_{,uv}+\Omega (u,v)\, f=0;\;\;\; \Omega (u,v)=-
{\left({\rm e}^{-U/2}\right)_{,uv}\over {\rm e}^{-U/2}}+{1\over 2}
{{\rm e}^{-M+U}}\left( k_x^2{\rm e}^{-V}+k_y^2{\rm e}^V\right).
\label{eq:f(u,v)}\ee
From now on, we will refer to $\Omega (u,v)$ as the {\em  potential term}.
In the flat region (region IV) this potential term is simply,

\bb \Omega _{\rm IV}(u,v)=L_1L_2(k_x^2+k_y^2), \label{eq:VIV}\ee
where we have used that the functions $U$ and $V$ in (\ref{eq:dsG})
are zero in the flat region, and the function $\exp (-M)=2L_1L_2$.
Using (\ref{eq:VIV}), equation (\ref{eq:f(u,v)}) can
be solved as,

\bb f(u,v)={\rm e}^{-i2{\hat k}_+u-i2{\hat k}_-v},\label{eq:fIV}\ee
where $k_\pm$ are two new separation constants with dimensions of
energy, and we define for convenience two dimensionless constants as
${\hat k}_\pm\equiv \sqrt{L_1L_2}\, k_\pm$. These new separation
constants are directly related to the previous ones $k_x$ and $k_y$
by,

\bb 4\, k_+k_-=k_x^2+k_y^2. \label{eq:k+-}\ee
The field solution in region IV reduces, thus, to the usual Minkowski
plane wave solution, i.e,

\bb \phi _k (u,v,x,y)={1\over\sqrt{2k_-(2\pi )^3}}\,
{\rm e}^{-i2{\hat k}_+u-i2{\hat k}_-v+ik_xx+ik_yy}. \label{eq:phiIV}\ee
These modes are well normalized on the null hypersurface
$\Sigma$, which is the boundary
of the plane wave regions II and III with the flat region IV. Even
though $\Sigma$ is a 
null hypersurface,
a well defined scalar product is given by (see 
\cite{dor93} for details),

\bb (\phi _1,\phi _2)=-i\int dx dy\left[
\int _{-\infty} ^0 \left.\left(\phi _1 
{\buildrel\leftrightarrow\over\partial}_u \phi
_2^*\right)\right|_{v=0}\, du +
\int _{-\infty} ^0 \left.\left(\phi _1 
{\buildrel\leftrightarrow\over\partial}_v \phi
_2^*\right)\right|_{u=0}\, dv\right].  \label{eq:scalarproduct}\ee
The modes (\ref{eq:phiIV}) will determine on $\Sigma$ a well 
posed set of boundary
conditions for modes in regions II and III. There, the
potential term in equation (\ref{eq:f(u,v)}) is simply,

\bb \Omega _{i}(u,v)={1\over 2}{\rm e}^{-M_{i}+U_{i}}
\left(k_x^2{\rm e}^{-V_{i}}+k_y^2{\rm e}^{V_{i}}\right), 
\label{eq:VII/III}\ee
where the label $i={\rm II}$ or $i={\rm III}$ in the functions 
$U$, $V$ and $M$, stands
for their particular values in the plane wave regions II or III.
Then, the solution of equation (\ref{eq:f(u,v)}) in regions II and III with
the boundary conditions imposed by the flat modes (\ref{eq:phiIV}) on
the hypersurface $\Sigma$, can be easily found as,

\bb f(u,v)=\left\{\begin{array}{ll}
{\rm e}^{-i2{\hat k}_-v-iA_{\rm II}(u)/(2{\hat k}_-)}; 
& {\rm in\; region\; II},\\
\\
{\rm e}^{-i2{\hat k}_+u-iA_{\rm III}(v)/(2{\hat k}_+)}; 
& {\rm in\; region\; III},
\end{array}\right. \label{eq:fII/III}\ee
where the generic function $A_i(\zeta )$, with $i=$II, III is given by,

\bb A_i(\zeta )=\int_0^\zeta  d\zeta '\, {1\over 2}
{\rm e}^{-M_i(\zeta ')+U_i(\zeta ')}
\left({\rm e}^{-V_i(\zeta ')}k_x^2+{\rm e}^{V_i(\zeta ')}k_y^2\right).
\label{eq:AII/III}\ee
Therefore, the well normalized ``in'' modes in regions II and III are,

\bb \phi (u,v,x,y)={1\over\sqrt{2k_-(2\pi)^3}}\,{\rm e}^{ik_xx+ik_yy}\,
\left\{\begin{array}{ll}
\displaystyle
{1\over\cos u}{\rm e}^{-i2{\hat k}_-v-iA^{}_{\rm II}(u)/(2{\hat k}_-)};
& {\rm in\; region\; II},\\
\\
\displaystyle
{1\over\cos v}{\rm e}^{-i2{\hat k}_+u-iA^{}_{\rm III}(v)/(2{\hat k}_+)};
& {\rm in\; region\; III},
\end{array}\right. \label{eq:phiII/III}\ee

Now the Cauchy problem is well posed on the boundary 
$\Sigma _ {\rm I}=\{u=0,\,0\leq v<1\}\cup\{0\leq u<1,\, v=0\}$ 
between plane wave regions II and III and
interaction region. Notice that the initial modes
(\ref{eq:phiII/III}) are well normalized on  the boundary $\Sigma$
between the flat region and the plane wave regions, and this means,
from general grounds, that they remain well normalized on the boundary
between the plane waves and the interaction region. This can be seen
explicitly using a well defined scalar product on the hypersurface
$\Sigma _{\rm I}$, which
similarly to (\ref{eq:scalarproduct}) is,

\bb (\phi _1,\phi _2)=-i\int dx dy\left[
\int _{0} ^{1}{\rm e}^{-U_{II}}(u) \left.\left(\phi _1 
{\buildrel\leftrightarrow\over\partial}_u \phi
_2^*\right)\right|_{v=0}\, du +
\int _{0} ^{1}{\rm e}^{-U_{III}}(v) \left.\left(\phi _1 
{\buildrel\leftrightarrow\over\partial}_v \phi
_2^*\right)\right|_{u=0}\, dv\right].  \label{eq:scalarproductI}\ee
The correct  normalization of modes (\ref{eq:phiII/III}) follows
easily from the
fact that functions $A_i(x)$ in (\ref{eq:AII/III}) are unbounded at
the folding singularities, ${\cal P}=\{u=1,\; v=0\}$ and ${\cal
P}'=\{u=0,\; v=1\}$.

We have now to solve equation (\ref{eq:f(u,v)}) in region I with the
boundary conditions imposed by (\ref{eq:phiII/III}) on the lines
$\Sigma _{\rm I}=\,\{u=0,\, 0\leq v<1\}\cup\{0\leq u<1,\, v=0\}$,
which are characteristic lines for the partial differential equation 
(\ref{eq:f(u,v)}). Thus, the only independent boundary
conditions, i.e. the Cauchy data, are the initial values of the
function $f(u,v)$ on them (the normal derivatives of
the field on the characteristics, which are usually part of the Cauchy
data, are determined by the values of the function $f(u,v)$ itself,
see \cite{gara64} for details).
Furthermore, if we recall that the collision center is determined by
the condition $u=v$, then 
the subset of Cauchy data that affects the neighborhood of the
collision center, i.e its causal past, lies on the segments,
${\hat\Sigma} _{\rm I}=\,\{u=0,\,0\leq v<{\hat s}\}\cup\{0\leq u<{\hat s},\,
v=0\}$ \cite{gara64}, where ${\hat s}^{n_1}+{\hat s}^{n_2}=1$
and $n_1$, $n_2$ are defined in (\ref{eq:p1p2}). 
We will denote the causal future region of these Cauchy data
(or equivalently, the causal past region of the collision center) by
region $\cal S$, see Fig. 2.

We can determine the behaviour of the solutions of equation
(\ref{eq:f(u,v)}) in the region near the
singularity using the following change of coordinates,

\bb \sigma  =b(v)+a(u),\;\;\; \rho =b(v)-a(u). \label{eq:sigrho}\ee
Then  equation (\ref{eq:f(u,v)}) becomes,

\bb f_{,\sigma\sigma}-f_{,\rho\rho}+\Omega (\sigma ,\rho)f=0,  \label{eq:fsigrho}\ee
where the potential term is,

\bb \Omega (\sigma ,\rho)={1\over 4 \sigma ^2}\left[1+
2{\rm e}^{-{\hat M}}{\sigma ^{\alpha -\beta +1}\over{\dot a}{\dot b }}
\left(
{\rm e}^{-{\hat V}}k_1^2+{\sigma ^{2\beta}}{\rm e}^{\hat V}k_2^2
\right)
\right], \label{eq:Vsigma}\ee
and where we have used for convenience two dimensionless separation
constants $k_1$ and $k_2$, directly related to $k_x$ and $k_y$ by,

\bb k_1=\sqrt{L_1L_2}\, k_x,\;\;\;\; k_2=\sqrt{L_1L_2}\, k_y.
\label{eq:kiL}\ee
Observe that this equation has a singular point at $\sigma =0$
due to the
term $\sigma ^{-2}$ on the potential $\Omega (\sigma ,\rho)$. In fact, we
have written the
functions $\exp (-M)$ and $\exp (\pm V)$ in the following way,

\bb {\rm e}^{-M}=\sigma ^{\alpha}{\rm e}^{-{\hat M}},\;\;\;
{\rm e}^{V}=\sigma ^{\beta}{\rm e}^{{\hat V}},\;\;\;
\alpha =(\beta ^2-1)/2,\;\;\;
\beta=p_1+p_2,
\label{eq:hatMV}\ee
as can be directly seen by (\ref{eq:Vuv})-(\ref{eq:Muv}),
where the new functions $\exp ({\hat M})$ and $\exp (\pm{\hat V})$,
are bounded at $\sigma =0$. It is straightforward to 
see from (\ref{eq:hatMV}) that $\alpha\pm\beta +1\geq 0$.
Thus, the singularity term in equation (\ref{eq:fsigrho}) is only
given by the factor $\sigma ^{-2}$.

Observe also that near the singularity
$\sigma =a(u)+b(v)=0$, the potential term (\ref{eq:Vsigma}) blows up
due to the factor $\sigma ^{-2}$ but it is perfectly smooth on 
coordinate $\rho =b(v)-a(u)$. Thus, the
relevant features are related to coordinate $\sigma$ only. Therefore,
we may not expect any physically remarkable difference 
if we substitute coordinate
$\rho$ in (\ref{eq:Vsigma}) by its constant value 
$\rho\equiv\rho _s=b({\hat s})-a({\hat s})$ when the 
singularity is approached from inside region $\cal S$, i.e. when
$u,\, v\rightarrow {\hat s}$ such that ${\hat s}^{n_1}+{\hat s}^{n_2}=1$.
Then, we will expect a solution
of equation (\ref{eq:fsigrho}) of the type 
$f(\sigma ,\rho )=g(\sigma )\exp (i k_3\rho )$, being $k_3$ a 
separation constant and where the new function
$g$ depends only on coordinate $\sigma$ and satisfies the differential
equation,

$$g_{,\sigma\sigma}+\left[\Omega (\sigma ,\rho _s)+k_3^2\right]\, g=0.$$
We could now attempt to solve this Schr\"odinger-type of differential equation.
For instance, we could try to find a solution by a WKB expansion.
Unfortunately, the initial conditions for equation (\ref{eq:fsigrho}) are
given on the boundaries $\Sigma _{\rm I}$ and there the  coordinates $\sigma$,
$\rho$ are not well defined. This means that, even though coordinates
$\sigma$ and $\rho$ would allow us to find a solution near the singularity
$\sigma =0$, they cannot be used to find a solution near the
boundaries with the plane waves, and therefore they are not convenient
for the mode propagation problem. Nevertheless, it will be
useful in what follows to know that this type of solution can be found
near the singularity. 

In fact, we should return to equation (\ref{eq:f(u,v)}) in coordinates $u$,
$v$ and we should recall that
we are only interested in finding the field solution in the   
neighborhood of the collision's center. Thus, the potential term of equation
(\ref{eq:f(u,v)}) is restricted into region $\cal S$, and the boundary
conditions are given on the segments ${\hat\Sigma}_{\rm I}$ (see Fig. 2). 
Then if,
in analogy with the change (\ref{eq:sigrho}),
we use the notation ${\sigma} ={\rm e}^{-U}$, such potential term is
simply given by,

\bb  \Omega ^{}(u,v)={{\dot a}{\dot b}\over 4\sigma ^2}+
{1\over 2}{\rm e}^{-{\hat M}}\sigma ^{\alpha -\beta -1}\left(
{\rm e}^{-{\hat V}}k_x^2+\sigma ^{2\beta}{\rm e}^{\hat V}k_y^2
\right), \label{eq:VIuv} 
\ee
with  $\alpha$, $\beta$ and 
${\rm e}^{-{\hat V}}$, ${\rm e}^{-{\hat M}}$ defined in
(\ref{eq:hatMV}). Observe that functions
${\rm e}^{-{\hat V}}$, ${\rm e}^{-{\hat M}}$ and ${\dot a}(u){\dot b}(v)$
are perfectly smooth in the entire region $\cal S$. However,
(\ref{eq:VIuv}) blows up when the term $\sigma =1-u^{n_1}-v^{n_2}$ 
goes to zero at the point $u=v={\hat s}$, where ${\hat s}^{n_1}+{\hat s}^{n_2}=1$. 
Taking all of this into account, it would be useful to find an 
adequate approximation for the potential (\ref{eq:VIuv}) such that:
i) it preserves the essential features of (\ref{eq:VIuv}), 
ii) it allows us to find a solution of equation (\ref{eq:f(u,v)})
which smoothly matches with the boundary conditions at
${\hat\Sigma}_{\rm I}$.

In order to find such an approximation it would be rather convenient
to introduce a new set of coordinates naturally adapted to this problem.
Recall that coordinates  (\ref{eq:sigrho}) are convenient to solve
equation  (\ref{eq:f(u,v)}) close to the singularity, but they are
not well defined globally and particularly on the boundaries with 
the single plane waves. Nevertheless, since $\sigma =a(u)+b(v)$
behaves near the singularity in region $\cal S$ as,
$\sigma\simeq {\dot a}_s\, (u-{\hat s})+{\dot b}_s\, (v-{\hat s})$,
with ${\dot a}_s={\dot a}({\hat s})$, ${\dot b}_s={\dot b}({\hat s})$, 
we may define
in analogy with (\ref{eq:sigrho}) a new time-like and  
space-like coordinates as,

\bb
t=\left[{\dot b}_s\, (v-{\hat s})+{\dot a}_s\, (u-{\hat s})\right]({\dot
a}_s{\dot b}_s)^{-1/2},\;\;\;\; 
z=\left[{\dot b}_s\, (v-{\hat s})-{\dot a}_s\, (u-{\hat s})\right]({\dot
a}_s{\dot b}_s)^{-1/2},
\label{eq:deftz}\ee
where in fact, ${t}$ and $z$ apply globally 
and close to the singularity in region $\cal
S$ they behave like coordinates (\ref{eq:sigrho}). Indeed,
$t\rightarrow\sigma ({\dot a}_s{\dot b}_s)^{-1/2}$ and
$z\rightarrow (\rho-\rho _s)({\dot a}_s{\dot b}_s)^{-1/2}$ as the
singularity $u=v={\hat s}$ is approached. 
The factor $({\dot a}_s{\dot b}_s)^{-1/2}$ is used to ensure that such a
coordinate change represents a Lorentz transformation.

We can easily find an expression for the function $\sigma (u,v)$ in
coordinates ${t}$, $z$ in the neighborhood of the singularity
$u=v={\hat s}$ by expanding $\sigma (u,v)$ in terms of a Taylor
series centered in $u=v={\hat s}$, with $u-{\hat s}$ and $v-{\hat s}$ 
expressed,
from (\ref{eq:deftz}), in terms of coordinates ${t}$ and $z$. The
result is,

\bb
\sigma ({t},z)=t\, ({\dot a}_s{\dot b}_s)^{1/2}
+\sum _{n=2}\left[
 {a_s^{(n)}\over {\dot a}_s^n}\left({{t}-z\over 2}\right)^n
+{b_s^{(n)}\over {\dot b}_s^n}\left({{t}+z\over 2}\right)^n
\right]\, { ({\dot a}_s{\dot b}_s)^{n/2}\over n!}.
\label{eq:sigtz}\ee
Observe, however, that in region $\cal S$ we expect the physical
remarkable effects to occur near the singularity $u=v={\hat s}$, where the
potential term (\ref{eq:VIuv}) grows unbounded as 
$\sigma ({t},z)\rightarrow 0$. Recall that in the neighborhood of
$\sigma ({t},z)=0$ the relevant term in (\ref{eq:sigtz}) is the
first order term $t\, ({\dot a}_s{\dot b}_s)^{1/2}$, which is
insensitive to coordinate $z$. Therefore, we may not expect any
physically remarkable difference if instead of working with the
function $\sigma ({t},z)$, which depends on both coordinates
${t}$ and z, we change to a new function $\sigma ({t})$
depending only on coordinate ${t}$ and directly derived from
(\ref{eq:sigtz}) by taking $z=0$ throughout the entire region
$\cal S$, i.e.,

\bb
\sigma ({t})=t\, ({\dot a}_s{\dot b}_s)^{1/2}
+\sum _{n=2}\left(
 {a_s^{(n)}\over {\dot a}_s^n}
+{b_s^{(n)}\over {\dot b}_s^n}
\right)\, \left({{t}\over 2}\right)^n \, 
{ ({\dot a}_s{\dot b}_s)^{n/2}\over n!}.
\label{eq:sigt}\ee
Observe also that the rest of the functions which appear in 
(\ref{eq:VIuv}), i.e. ${\rm e}^{\pm{\hat V}(u,v)}$, 
${\rm e}^{-{\hat M}(u,v)}$ and ${\dot a}(u){\dot b}(v)$,
are perfectly smooth all over region $\cal S$. Thus, we may not also
expect any physically remarkable difference if we substitute these
functions, which also depend on both coordinates ${t}$ and $z$,
by approximate functions depending on coordinate ${t}$ alone.
The procedure would be equivalent to the one used for
$\sigma (u,v)$. Namely, given a smooth function $\psi (u,v)$ in region
$\cal S$, we expand it by a Taylor series centered in the singularity
$u=v={\hat s}$. Then, we substitute $u-{\hat s}$, $v-{\hat s}$ by coordinates
${t}$, $z$ using the definition (\ref{eq:deftz}) and finally we take
$z=0$. The result would simply be,

\bb
\psi ({t})=\psi ({\hat s},{\hat s})
+\sum _{n=1}\left.\left(
 {1\over {\dot a}_s}{\partial\over\partial u}
+{1\over {\dot b}_s}{\partial\over\partial v}
\right)^n \psi (u,v)\right|_{u=v={\hat s}} 
\left({{t}\over 2}\right)^n \,
{ ({\dot a}_s{\dot b}_s)^{n/2}\over n!}.
\label{eq:psit}\ee

Recall that the boundary conditions for equation (\ref{eq:f(u,v)})
restricted in region $\cal S$, lie on the segments ${\hat\Sigma}_{\rm I}$ and
since the endpoints of these segments are not close to the
singularities $\cal P$ or ${\cal P}'$, we may suppose that the boundary
conditions are not significantly different from the flat boundary
conditions.
This fact, together with the prior replacements of $u, v$-dependent 
functions by related ${t}$-dependent functions
in the potential term (\ref{eq:VIuv}), suggest that we may reproduce
the main physical features if we replace the four-region mode
propagation problem by a much simpler problem. 
Such a problem would consist in solving equation (\ref{eq:f(u,v)})
with a potential term (\ref{eq:VIuv}) depending on coordinate ${t}$ alone
and where the  boundary conditions would be imposed
by the flat Minkowski modes (\ref{eq:phiIV}) below the hypersurface
$\{{T}=0,\; -{\hat s}<Z<{\hat s}\}$, where we denote
$T=u+v$, $Z=v-u$ (see Fig. 3).
Recall, however, that all this discussion is absolutely non
applicable when 
a solution for equation (\ref{eq:f(u,v)}) in the neighborhood of the
folding singularities $\cal P$ or ${\cal P}'$ is required. This is non
applicable because the boundary conditions (\ref{eq:phiII/III}) are
unbounded as the folding singularities are approached and therefore
these boundary conditions strongly differ from their counterparts
(\ref{eq:phiIV}) in the flat region.

In order to solve this Schr\"odinger-type problem, rather than relying
on the discussed approximations for the exact field equation
(\ref{eq:f(u,v)}), we will rewrite a new field equation using an
adequate approximation for the line element throughout the entire
causal past of the collision center.
This new approach, which  may seem redundant in the case of
the {\em particle production problem}, is absolutely necessary when
the renormalization of the stress-energy tensor is
discussed. This is essentially because the process of renormalization involves
the subtraction of the infinite divergences that arise from the
formal definition of the stress-energy tensor, and these divergences
can be expressed as entirely geometric terms, which are independent
of any possible 
approximations in the field equation. This means that in order to
recover the geometric divergences in the stress-energy tensor,  any
approximation in the field equation must be 
related to a suitable  approximation in the space-time geometry.

The approximations just discussed above can be essentially
recovered by
changing the exact colliding plane wave line element throughout the
causal region of the collision's center (\ref{eq:dsG}) into a related
line element. This new line element would basically 
consist in  replacing the metric coefficients
$\sigma (u,v)={\rm e}^{-U(u,v)}$, $\sigma (u,v)^{\pm\beta}{\rm
e}^{\pm{\hat V}(u,v)}$,  
$\sigma (u,v)^{-\alpha}{\rm e}^{-{\hat M}(u,v)}$ in (\ref{eq:dsG}), which
depend on both coordinates $u$ and $v$,  by new functions of coordinate
${t}$ alone in the sense described in (\ref{eq:sigt})-(\ref{eq:psit}), i.e.,

\bb
d{\hat s}^2_{\rm I}={1\over 2}\,\sigma({t})^\alpha\,{\rm e}^{-{\hat M}({t})}\,
\left(d{t} ^2-d{z} ^2\right)
-\sigma({t})^{1+\beta}\left(
{\rm e}^{{\hat V}({t})}\, dx^2
+\sigma({t})^{-2\beta}\,{\rm e}^{-{\hat V}({t})}\, dy^2
\right).
\label{eq:dshatIb}
\ee
We will suppose that the line element (\ref{eq:dshatIb})
applies all over the causal past of the collision center, not
only in the interaction region but also through the plane wave regions
II and III in the sense of Fig. 3.
The plane wave collision starts at 
$t=t_0=-{\hat s}\,({\dot b}_s+{\dot a}_s)\, ({\dot a}_s{\dot b}_s)^{-1/2}$ but
to avoid smoothness problems derived from such an
approximation, we will suppose that (\ref{eq:dshatIb}) applies
exactly on a range $t_0+\epsilon <{t}<0$, for a certain
$\epsilon >0$. In the range
$t_0\leq {t}\leq t_0+\epsilon$, as described
below,  we will interpolate a line element 
which smoothly matches with the flat space at $t=t_0$.
Nevertheless, the particular details of this matching 
will not affect the main physical features.

The
exact field equation for this approximate space-time is,

\bb
\left(\Box +\xi  R\right)\phi =0,
\label{eq:Appfieldeq}\ee
where it is necessary to consider a coupling curvature term in the field
equation because, although the exact space-time is a vacuum
solution, we have a bounded nonzero value for
$R$ in the approximated space-time.
In order to solve this new field equation, we start rewriting  the line
element (\ref{eq:dshatIb}) in the following general way,

\bb
ds^2=(f_1f_2f_3)\, d{{t} ^*}^2-\left(f_1f_2\over f_3\right)\, d{z} ^2
-\left(f_2f_3\over f_1\right)\, dx^2-
\left(f_1f_3\over f_2\right)\, dy^2,
\label{eq:dsgen}
\ee
where the $f_i$ are functions of coordinate ${t}$ alone, which
for values of $t_0+\epsilon <{t}<0$, can be
straightforwardly determined by direct comparison with
(\ref{eq:dshatIb}) as 

\bb
f^2_1(t)={1\over 2}{\rm e}^{-{\hat V}(t)-{\hat M}(t)}\,
\sigma (t)^{(\beta -1)^2/2},\;\;\;\;
f^2_2(t)={1\over 2}{\rm e}^{ {\hat V}(t)-{\hat M}(t)}\,
\sigma (t)^{(\beta +1)^2/2},\;\;\;\;
f^2_3(t)=\sigma ^2(t),
\label{eq:dsgentf}
\ee
where definitions (\ref{eq:hatMV}) have been used.
For values ${t}\leq t_0$ we take $f_1({t})=f_2({t})=\sqrt{L_1L_2}$, 
$f_3({t})=1$, which correspond to their values in flat
space. Finally,  in
the interval $t_0\leq{t}\leq t_0+\epsilon$, we smoothly interpolate each  
$f_i({t})$
($i=1,2,3$) between these values.
Also,
in order to prevent singularities in the field
equation, we conveniently reparametrize coordinate ${t}$, by ${t}
^*({t})$, as follows,

\bb
{d{t} ^*\over d{t}}={1\over f_3({t})}.
\label{eq:dxi*dxi}\ee
Now, we use the following ansatz for the field solutions,

\bb
\phi _k=h({t} ^*)\,{\rm e}^{ik_xx+ik_yy+ik_{z}{z}},
\label{eq:ansatz}\ee
where the plane wave factor in coordinates $x$, $y$ is related to the
translational symmetry of the space-time along  the transversal directions
$x$, $y$, and the plane wave factor in
coordinate ${z}$ is just a consequence of our approximation. 
Then equation (\ref{eq:Appfieldeq})
directly leads to the following Schr\"odinger-like differential
equation for the function $h({t} ^*)$,

\bb
h_{,{t} ^*{t} ^*}+\omega ^2({t} )\, h=0,\;\;\;\;
V({t})\equiv\omega ^2({t} )=f_0^2({t} )+f_1^2({t} )\, 
k_x^2+f_2^2({t} )\, k_y^2+
f_3^2({t} )\, k_z^2,
\label{eq:heq}\ee
where the function $f_0({t})$ stands for,

\bb
f_0^2({t} )=\left[f_1({t} )f_2({t} )f_3({t} )\right]\,\xi  R.
\label{eq:omegafi}\ee
Observe from (\ref{eq:p1p2}) and (\ref{eq:hatMV}), 
that the range of the parameter $\beta$ in (\ref{eq:dsgentf})
is $\beta ^2 <8$. Then,
we always have that either $(\beta +1)^2/4<1$ or $(\beta -1)^2/4<1$.

Now, it would be rather convenient to discuss 
whether or not it makes sense to solve the 
differential equation (\ref{eq:heq}) in terms of a
WKB expansion. Observe that, when the parameter $\beta =\pm 1$ in
(\ref{eq:dsgentf}), then the frequency $\omega ({t})$ goes to
a non-zero constant value towards the singularity $\sigma =0$.
Then the short-wavelength condition, i.e. $\omega
^{-1}d/d{t}^*\ln\omega\ll 1$, holds and is particularly accurate
near the singularity. Hence, equation (\ref{eq:heq}) admits
everywhere  a WKB solution.

However, when $\beta\neq\pm 1$, the exponents of the $\sigma$-terms
in the functions $f_i$ in (\ref{eq:dsgentf}) are all non-zero. 
Thus, all the functions $f_i$
vanish at the singularity $\sigma =0$.
Also the function
$f_0^2=\xi\,R\,(f_1f_2f_3)$ vanishes at $\sigma =0$. This can
be easily seen because  $f_0$ can be expressed
in terms of $f_1$, $f_2$ and $f_3$, as

$$
f_0^2=\xi\,\left[
{{\ddot f}_1\over f_1}+{{\ddot f}_2\over f_2}+{{\ddot f}_3\over f_3}
-{1\over 2}\,\left(
{{\dot f}_1\over f_1}+{{\dot f}_2\over f_2}+{{\dot f}_3\over f_3}
\right)^2
\right],
$$
and substituting the values (\ref{eq:dsgentf}) we have that
$f_0^2$ is proportional to $\sigma$. Therefore,
the frequency term $\omega ({t})$ in (\ref{eq:heq}) goes to
zero towards the singularity $\sigma =0$. Hence, the short wavelength
condition does not hold in the region close to
the singularity and thus a WKB solution for equation (\ref{eq:heq})
is not directly appropriate in this region. Nevertheless, we can still
use a WKB solution, which at least would be appropriate if we are not
close enough to the singularity $\sigma =0$ (recall that $\sigma =0$
means from (\ref{eq:dxi*dxi}) that ${t}^*\rightarrow\infty$). 
Observe, however, that such a WKB expansion may still give
information near the singularity. The reason lies upon the fact that we 
can interpret this WKB solution as a limit case of a set of WKB
expansions which are truly appropriate near the singularity. Such a 
limit process may be briefly introduced as follows:
we add a small quantity to $V({t})$ in (\ref{eq:heq}).
Then, the short wavelength holds and  it reduces to
$(d{t}/d{t}^*)\, dV/d{t}\ll 2\,\omega ^3$, where now the frequency
$\omega ({t})$ is non-zero and 
$d{t}/d{t}^*=f_3({t})\rightarrow 0$ towards the singularity.
Hence,  the differential equation (\ref{eq:heq}) 
admits a WKB solution everywhere. We could then perform the
involved calculations and finally set the small quantity that we have
introduced in $V({t})$ to zero. Mathematically, adding this small quantity
to $V({t})$ would be equivalent to subtract a 
small amount $\delta >0$ into the exponents of 
the $\sigma$-terms of functions
$f_1$ and $f_2$ in (\ref{eq:dsgentf}), i.e.,  

\bb
f^2_1(t)={1\over 2}{\rm e}^{-{\hat V}(t)-{\hat M}(t)}\,
\sigma (t)^{(\beta -1)^2/2-2\delta},\;\;\;\;
f^2_2(t)={1\over 2}{\rm e}^{ {\hat V}(t)-{\hat M}(t)}\,
\sigma (t)^{(\beta +1)^2/2-2\delta},
\label{eq:f1f2lamb}
\ee
which is also equivalent to slightly change
the line element (\ref{eq:dsgen}) by including a factor
$\sigma ({t})^{-2\delta}$ to the metric coefficients $g_{{t}^*{t}^*}$,
$g_{zz}$ only. Then, the function $f_0$ is finite at the singularity and
is given by $f_0^2=2{\dot a}_s{\dot b}_s\delta + O(\sigma )$.

Finally, using a WKB expansion, the mode solutions
$\phi _k$ which reduce to the flat mode solutions in the region prior
to the arrival of the waves, are

\bb
\phi _k={{\hat\omega}^{1/2}\over\sqrt{(2\pi)^32k_-W({t} )}}
{\rm e}^{ik_xx+ik_yy+ik_3{z}-i\int ^{{t} ^*}W(\zeta )d\zeta ^*},
\label{eq:phifi}\ee
where we denote ${\hat\omega}^2=k_1^2+k_2^2+k_3^2$
with $k_1$ and $k_2$ given in (\ref{eq:kiL}),
$k_3=k_z$ and where $W({t} )$ stands for an
adiabatic series in powers of the time-dependent frequency 
$\omega ({t} )$ of the 
modes and its derivatives. Up to adiabatic order four (i.e. up
to terms involving four derivatives of $\omega ({t})$) 
$W({t})$ it is given by,

\bb W({t} )=\omega  +{A_2\over\omega ^3} +{B_2\over\omega ^5} 
+{A_4\over\omega ^5}
 +{B_4\over\omega ^7} +{C_4\over\omega ^9} +{D_4\over\omega 
^{11}},\label{eq:Wxi}\ee
where, using the notation ${\dot V}\equiv dV/d{t}^*$,
 
\bb A_2=-{{\ddot V}\over 8},\;\;\;\; B_2={5\over 32}\,{{\dot 
V}^2},\label{eq:An}\ee
 
$$ A_4={{\ddddotV}\over 32},\;\;\;\; B_4=-
{28\,{\dot V}\,{\dddotV}+19\,{\ddot
V}^2\over 128},\;\;\;\; C_4={221\over 258}\,{{\dot V}^2\,{\ddot V}},\;\;\;\;
D_4=-{1105\over
2048}\,{{\dot V}^4},$$
and 
$A_n$, $B_n$, ... denote the $n$ adiabatic terms in
$W({t})$.
Up to adiabatic order zero it is simply
$W({t} )=\omega({t} )$. Observe the two following facts:

\noindent
(i) Near the singularity $\sigma =0$ we have $W({t})\simeq\omega({t})$.
This is because the higher adiabatic corrections vanish at the singularity.
In fact, for the case $\beta =\pm 1$ in (\ref{eq:dsgentf}), these
higher adiabatic corrections naturally vanish. Also, for the
case $\beta\neq\pm 1$ and under the limit prescription introduced in
(\ref{eq:f1f2lamb}), they also vanish.

\noindent
(ii) In the flat region prior to the arrival of the waves we have
$W({t} )={\hat\omega} =\left(k_1^2+k_2^2+k_3^2\right)^{1/2}$. In that case,
since $f_3=1$, we can use (\ref{eq:dxi*dxi}) to set
$ {t} ^*={t}$, where without loss of generality we choose the
value ${t}^*=t_0$ at ${t=t_0}$.
Therefore, the mode solutions (\ref{eq:phifi}) in the flat region reduce to,

\bb
\phi _k^{{\rm IV}}={1\over\sqrt{(2\pi)^32k_-}}{\rm
e}^{ik_xx+ik_yy+ik_{z}{z} -i{\hat\omega}{t}}.
\label{eq:phiIVL}\ee
To see how (\ref{eq:phiIVL}) are related to the flat modes
(\ref{eq:phiIV}), recall first that the new separation constant
$k_{z}$ is related to the
original $k_\pm$ by the ordinary null momentum relations, i.e.,

\bb
{\hat\omega}={\hat k}_++{\hat k}_-,\;\;{k_{z}}={\hat k}_+-{\hat k}_-.
\label{eq:nullmom}\ee
In fact, flat modes (\ref{eq:phiIV}) have the same formal expression
as (\ref{eq:phiIVL}) where instead of $t$ there is a
time-like coordinate $T=v+u$ and instead of $z$ there is
a  space-like coordinate $Z=v-u$,
and coordinates $(t,\, z)$ and $(T,\, Z)$ are simply related by a
Lorentz boost such that, $t=\gamma (T+\beta Z)$, $z=\gamma (Z+\beta
T)$, with 

\bb
\gamma ={{\dot b}_s+{\dot a}_s\over 2\sqrt{{\dot a}_s{\dot
b}_s}},\;\;\;\;
\beta ={{\dot b}_s-{\dot a}_s\over {\dot b}_s+{\dot a}_s}.
\label{eq:gammabeta}\ee
Unfortunately, we use the same notation $\beta$ in (\ref{eq:gammabeta}) and in
(\ref{eq:hatMV}), but note that there is not possible confusion.

It is important to understand, however, that we are constructing a set of mode
solutions as an adiabatic series in terms of derivatives of the
frequency $\omega ({t})$ in the differential equation
(\ref{eq:heq}). This procedure is similar but not equivalent to the
construction of an {\em adiabatic vacuum state} where the field modes
are expanded as an adiabatic series in terms of the derivatives of the
metric coefficients (see for example \cite{bir82} for details). In
fact, observe for instance that the term $f^2_0({t})$ in (\ref{eq:heq})
involves two derivatives of the metric since it is directly related to
the curvature scalar. Thus, it would be a second order term for 
an eventual adiabatic vacuum, but it is simply a zeroth order term in 
our adiabatic series in derivatives of $\omega ({t})$.

\section{Hadamard function in the interaction region}

The key ingredient to calculate the vacuum expectation value of the
stress-energy tensor is the {\em Hadamard function} $G^{(1)}(x,x')$,
which is defined as the vacuum expectation value of the anticommutator
of the field, i.e.,

\bb G^{(1)}(x,x')=\langle\{\phi (x),\phi (x')\}\rangle =\sum _k 
\left\{u_{k}(x)\, u^*_{k}(x')+u_{k}(x')\,
u^*_{k}(x)\right\}. \label{eq:defHadamard}\ee
This Hadamard function contains non-physical divergence terms which
can be subtracted by the following point splitting prescription,

\bb G_B^{(1)}(x,x')=G^{(1)}(x,x')-S(x,x'),
\label{eq:GB}\ee
where $S(x,x')$ is the {\em midpoint expansion} of a locally 
constructed quantity commonly referred as
a {\em Hadamard elementary solution} 
(see for example \cite{wal94}) and given by

\bba
S(x,x')&=&{1\over 8\pi ^2}\left\{
-{2\over\sigma}
-2{\Delta ^{(2)}}_{{\bar\mu}{\bar\nu}}\,{\sigma ^{\bar\mu}\sigma
^{\bar\nu}\over\sigma}
-2{\Delta ^{(4)}}_{{\bar\mu}{\bar\nu}{\bar\rho}{\bar\tau}}
\,{\sigma ^{\bar\mu}\sigma ^{\bar\nu}\sigma ^{\bar\rho}\sigma
^{\bar\tau}\over\sigma}
 -a_1^{(0)}\,\ln (\mu ^{-2}\sigma ) 
\right.\nonumber
\\
& & \left.
-\left[
\left(
a_1^{(0)}{\Delta ^{(2)}}_{{\bar\mu}{\bar\nu}}
+{a_1 ^{(2)}}_{{\bar\mu}{\bar\nu}}
\right)\sigma ^{\bar\mu}\sigma ^{\bar\nu}
-{1\over 2}a_2^{(0)}\sigma
\right]\,\ln (\mu ^{-2}\sigma )
-{3\over 4}a_2^{(0)}\sigma
\right\},
\label{eq:Hada1}\eea
where the coefficients ${\Delta ^{(2)}}_{{\bar\mu}{\bar\nu}}$,
${\Delta ^{(4)}}_{{\bar\mu}{\bar\nu}{\bar\rho}{\bar\tau}}$,
$a_1^{(0)}\cdots$ are written
in Appendix A. We use the standard definition for the {\em geodetic
biscalar} $\sigma (x,x')=(1/2)s^2(x,x')$, being $s(x,x')$ the proper
distance between the points $x$ and $x'$ on a non-null geodesic
connecting them. Also, 
$\sigma _{\bar\mu} (x,x') =(\partial /\partial x^{\bar\mu})\sigma (x,x')$ 
is a geodesic tangent vector at the point $\bar x$ with modulus $s(x,x')$,
being $\bar x$ the {\em midpoint} between $x$ and $x'$ on the geodesic.
The parameter $\mu$ in the
logarithmic term of (\ref{eq:Hada1}) is an arbitrary length parameter,
which is related to the two-parameter ambiguity of the point-splitting 
regularization scheme \cite{wal94}.
Then we can compute $\langle T_{\mu\nu} \rangle$ by means of the
following differential operation,

\bb \langle T_{\mu\nu}(x)\rangle =\lim _{x\rightarrow x'}\, {\cal 
D}_{\mu\nu}G^{(1)}(x,x'), \label{eq:limDT}\ee
where ${\cal D}_{\mu\nu}$ is a nonlocal differential operator, which
in the conformal coupling case ($\xi =1/6$) is given by,

\bba{\cal D}_{\mu\nu}&=&
{1\over 6}\,\left(\nabla _{\mu '}\nabla _{\nu}+\nabla
_{\nu'}\nabla_{\mu}\right)
-{1\over 24}\, g_{\mu\nu}\,\left(
\nabla _{\alpha '}\nabla ^{\alpha}+
\nabla _{\alpha }\nabla ^{\alpha '}\right)-\nonumber\\
& &{1\over 12}\,\left(\nabla _{\mu }\nabla _{\nu}+\nabla
_{\mu'}\nabla_{\nu '}\right)
+ {1\over 48}\, g_{\mu\nu}\,\left(
\nabla _{\alpha  }\nabla ^{\alpha  }+
\nabla _{\alpha '}\nabla ^{\alpha '}\right)-\nonumber\\
& &{1\over 12}\,\left(R_{\mu\nu}-{1\over 4}R\, g_{\mu\nu}\right).
\label{eq:Dopdif}\eea
However, the above differential operation and its limit have no immediate 
covariant meaning because
$G^{(1)}(x,x')$ is not an ordinary function but a {\it biscalar} and the 
differential operator ${\cal
D}_{\mu\nu}$ is {\it nonlocal}; thus we need to deal with the nonlocal 
formalism of {\it bitensors} (see, for example \cite{dew60,chr76} or the
Appendix B of reference \cite{dor96} for a review on
this subject).

The regularization procedure (\ref{eq:GB}), however, 
fails to give a covariantly conserved 
stress-energy tensor essentially because the locally constructed
Hadamard function (\ref{eq:Hada1}) is not in
general symmetric on the endpoints $x$ and $x'$ (i.e. it satisfies the
field equation at the point $x$ but fails to satisfy it at $x'$)
(see \cite{wal78} for details).
Thus, to ensure covariant conservation, we must introduce an additional 
prescription:
 
\bb \langle T_{\mu\nu}(x)\rangle =\langle T_{\mu\nu}^B (x)\rangle
- {a^{(0)}_2(x)\over 64\pi ^2}\, g_{\mu\nu}. \label{eq:GBT}\ee
Note that this last term is responsible for the trace anomaly in the conformal 
coupling case, because even though
$\langle T_{\mu\nu}^B (x)\rangle$ has null trace when $\xi =1/6$, the trace of 
$\langle T_{\mu\nu}(x)\rangle$
is given by
$\langle T^{\mu}_{\mu}\rangle =
- {a^{(0)}_2(x)/(16\pi ^2)}$.
The regularization prescription just given in (\ref{eq:GBT})
satisfies the well known four  
Wald's axioms
\cite{wal94,wal76,wal77-78b,chr75}, a set of properties that any physically 
reasonable expectation value of the
stress-energy tensor of a quantum field should satisfy. There is still an 
ambiguity in this prescription since
two independent conserved local curvature terms, which are quadratic in the 
curvature, can be added to this
stress-energy tensor. In particular, the $\mu$-parameter ambiguity in 
(\ref{eq:Hada1}) is a consequence of this (see \cite{wal94} for details).
Such a two-parameter ambiguity, however, cannot be 
resolved within the limits of the
semiclassical theory, it may be resolved in a complete quantum theory
of gravity 
\cite{wal94}. Note, however, that in some sense this ambiguity does not affect 
the knowledge of the matter
distribution because a tensor of this kind belongs properly to the left hand 
side of Einstein equations, i.e.
to the geometry rather than to the matter distribution.

\section{Procedure to compute $\langle T_{\mu\nu}\rangle$}

After this preliminary introduction on the point-splitting regularization
technique, we may proceed to  calculate the Hadamard function $G^{(1)}(x,x')$ 
in the interaction region for the initial vacuum state
defined by the modes $\phi _k$, (\ref{eq:phifi}). The 
Hadamard function can be written as,
 
\bb G^{(1)}(x,x')=\sum _k\phi
_k(x)\,\phi ^*_k(x')\; +{c.c.}    
\label{eq:G(1)}\ee
Note that solutions $\phi _k$ contain the function $h({t} ^*)$, which
cannot be calculated analytically but may be approximated
up to any adiabatic order as described in (\ref{eq:Wxi})-(\ref{eq:An}). 
Thus, we have the 
inherent ambiguity of where to
cut the adiabatic series. In fact, this is an asymptotic expansion,
which has a well established ultraviolet limit but it may have convergence
problems in the low-energy limit.
However, observe from (\ref{eq:dxi*dxi}) and (\ref{eq:heq})
that since: (i) $dt/dt^*\rightarrow 0$, 
(ii) $V({t})=\omega ^2({t})$ is bounded and, under the
prescription introduced in (\ref{eq:f1f2lamb}), is also non-zero
towards the singularity, then the adiabatic series (\ref{eq:Wxi})
reduces to $W\simeq\omega$ near the singularity. 
This means that we could cut the 
adiabatic series (\ref{eq:Wxi}) at order zero if we were
interested in a calculation near the singularity. However, this is only
partially true. In fact, it would be true if we were only interested in the 
{\em particle production problem}, which essentially would involve
the evaluation of a Bogoliubov transformation between two different sets
of field modes. Unfortunately it is not
sufficient for the calculation of the vacuum expectation value of the
stress-energy tensor.
This is because
$G^{(1)}$ calculated with $h({t}^*)$
at order zero does not reproduce the short-distance singular 
structure of a Hadamard elementary
solution (\ref{eq:Hada1}) in the coincidence limit $x\rightarrow x'$. The 
smallest adiabatic
order for the function $h({t}^*)$ 
which we need to recover the singular structure of $G^{(1)}$ is order 
four, basically  because our adiabatic construction of the mode
solutions is similar (but not equivalent) to an {\em adiabatic vacuum
state} (see \cite{bir82} for details).
 
In the mode sum (\ref{eq:G(1)}) we use the shortened notation 
$\sum 
_k\equiv 
\int ^{\infty}_{0}{dk_-/ k_-}\,
\int ^{\infty}_{-\infty}dk_x\,
\int ^{\infty}_{-\infty}dk_y$ or equivalently
$\sum _k\equiv
(L_1L_2)^{-1}
\int ^{\infty}_{-\infty}dk_1\,
\int ^{\infty}_{-\infty}dk_2\,
\int ^{\infty}_{-\infty}{dk_3/ {\hat\omega}}$,
where the change of variables
(\ref{eq:nullmom}) and the usual notation (\ref{eq:kiL}) have been used.
Therefore we have, 
            
\bba  G^{(1)}(x,x')&=& {1\over 2(2\pi)^3\, L_1L_2}\,
\int ^{\infty}_{-\infty}\int ^{\infty}_{-\infty}\int
^{\infty}_{-\infty}
{dk_1\, dk_2\, dk_3\over \sqrt{W({t})W({t} ')}}\,\times \nonumber\\
& &{\rm e}^{-i\int _{{t^*} '}^{{t^*}}W({\zeta})d{\zeta}^*+ik_x
(x-x')+ik_y (y-y')+ 
ik_{z} ({z} -{z} ')}\;\; +c.c.                         
\label{eq:G(1)2}\eea
We assume that the points 
$x$ and $x'$ are connected by a
non-null geodesic in such a way that they are at the same proper distance 
$\epsilon$ from a third midpoint
$\bar x$. We parametrize the geodesic by its proper distance $\tau$ and 
with abuse of notation we denote 
the end points by
$x$ and $x'$, which should not be confused with the third
component of $({t} ,\;{z} ,\; x,\; y)$. Then we expand the integrand
function in powers of $\epsilon$ and we finally integrate term by term
to get an expression up to $\epsilon ^2$. The details of such a tedious
calculation can be found in ref. \cite{dor97}. The result is,
 
\bba G^{(1)}(x,x')&=&{\bar A}+\sigma\,{\bar b}
+{C}_{{\bar \alpha}{\bar \beta}}\,{\sigma}^{\bar \alpha}{\sigma}^{\bar\beta}
+{D}_{{\bar \alpha}{\bar \beta}{\bar \gamma}{\bar \delta}}\,
{\sigma}^{\bar \alpha}{\sigma}^{\bar\beta}{\sigma}^{\bar
\gamma}{\sigma}^{\bar\delta}
+{1\over 8\pi ^2}\left\{
-{2\over\sigma}
-2{\Delta ^{(2)}}_{{\bar\mu}{\bar\nu}}\,{\sigma ^{\bar\mu}\sigma
^{\bar\nu}\over\sigma}
\right.
\label{eq:Hada2}\\
& & \left.
-2{\Delta ^{(4)}}_{{\bar\mu}{\bar\nu}{\bar\rho}{\bar\tau}}
\,{\sigma ^{\bar\mu}\sigma ^{\bar\nu}\sigma ^{\bar\rho}\sigma
^{\bar\tau}\over\sigma}
-a_1^{(0)}\, L
-\left[
\left(
a_1^{(0)}{\Delta ^{(2)}}_{{\bar\mu}{\bar\nu}}
+{a_1 ^{(2)}}_{{\bar\mu}{\bar\nu}}
\right)\sigma ^{\bar\mu}\sigma ^{\bar\nu}
-{1\over 2}a_2^{(0)}\sigma
\right]\, L
\right\}
\nonumber\eea
where $L$ is a logarithmic term defined as 
$L=2\gamma+\ln (\sigma\,\xi R/2)$ and
$\gamma$ is the Euler's constant. All the involved
coefficients
$\bar A$, $\bar b$, ${C}_{{\bar \alpha}{\bar \beta}}$... depend on which
particular space-time, belonging to the
Szekeres Class of solutions, we are performing the calculations.
 
According to (\ref{eq:GB}), the Hadamard function can be regularized using the 
elementary
Hadamard solution (\ref{eq:Hada1}) and finally the regularized expression for
$G^{(1)}(x,x')$ up to order $\epsilon ^2$ is,
 
\bba G^{(1)}_B(x,x')&=&{\bar A}+\sigma\,{\bar B}
+{C}_{{\bar \alpha}{\bar \beta}}\,{\sigma}^{\bar \alpha}{\sigma}^{\bar \beta}
+{D}_{{\bar \alpha}{\bar \beta}{\bar \gamma}{\bar \delta}}\,
{\sigma}^{\bar \alpha}{\sigma}^{\bar\beta}{\sigma}^{\bar
\gamma}{\sigma}^{\bar\delta}
\nonumber\\
& &+{1\over 8\pi ^2}\left\{
-a_1^{(0)}\, {\hat L}
-\left[
\left(
a_1^{(0)}{\Delta ^{(2)}}_{{\bar\mu}{\bar\nu}}
+{a_1 ^{(2)}}_{{\bar\mu}{\bar\nu}}
\right)\sigma ^{\bar\mu}\sigma ^{\bar\nu}
-{1\over 2}a_2^{(0)}\sigma
\right]\, {\hat L}
\right\},
\label{eq:G(1)fR}\eea
where ${\hat L}$ is a bounded logarithmic term given by,
${\hat L}=2\gamma+\ln (\mu ^{2}\,\xi R/2)$, $\mu$ being the arbitrary 
length parameter introduced in (\ref{eq:Hada1}),
and where the coefficient $\bar B$ is 
${\bar B}=b+3\, a_2^{(0)}/(32\pi ^2)$.
From (\ref{eq:G(1)fR}) we can directly read off the regularized mean
square field in  
the ``in" vacuum state as
$\langle\phi ^2\rangle ={\bar A}/2-a_1^{(0)}{\hat L}/(16\pi ^2)$.
It is important to remark, however, that the term 
${D}_{{\bar \alpha}{\bar \beta}{\bar \gamma}{\bar \delta}}$ in
(\ref{eq:G(1)fR}) 
appears only as a consequence of our approximate procedure of
calculating the Hadamard function, i.e.  using an adiabatic order four
expansion for the initial modes in powers of the 
mode frequency $\omega ({t})$ and its derivatives.
Had we used an exact expression for the
initial modes (or an adiabatic vacuum state \cite{bir82}), such
a term would not appear. 
 
Now, to calculate the vacuum expectation value of the stress-energy
tensor we have to apply the differential operator 
(\ref{eq:Dopdif}) to
(\ref{eq:G(1)fR}). As we have already pointed out, this is not straightforward 
because we
work with nonlocal quantities. Note first that the operator (\ref{eq:Dopdif}) 
acts on bitensors which depend on the end points
$x$ and $x'$, but the expression (\ref{eq:G(1)fR}) for $G^{(1)}_B$ depends on 
the midpoint $\bar x$. This means
that we need to covariantly expand (\ref{eq:G(1)fR}) in terms of 
the endpoints $x$ and $x'$. Also, the presence of quartic $\sigma ^\mu$
terms in (\ref{eq:G(1)fR}) gives, after differentiation, path dependent
terms which must be conveniently averaged.
The details of such a calculation
may be found for instance in \cite{dor96,dor97}.
Then, in
the orthonormal basis 
$\theta _1={g^{1/2}_{{t}{t}}}\,d{t}$,
$\theta _2={g^{1/2}_{{z}{z}}}\, d{z}$, $\theta _3={g^{1/2}_{xx}}\, dx$, 
$\theta _4={g^{1/2}_{yy}}\, dy$,  using the trace anomaly
prescription (\ref{eq:GBT}),  we may obtain
the expectation values $\langle T_{\mu\nu}\rangle$ in
the conformal coupling case and for values
$t_0+\epsilon<{t}<0$ of coordinate ${t}$.
For values of $t\leq t_0$, $\langle T_{\mu\nu}\rangle =0$, 
and to be consistent with the approximation 
we have used for the space-time geometry, we should
require that the value of $\langle T_{\mu\nu}\rangle$  goes 
smoothly to zero as ${t}\rightarrow t_0$. In fact, this can be achieved
using an adequate matching of the line element (\ref{eq:dshatIb}) with the
flat line element through the interval $t_0\leq{t}\leq t_0+\epsilon$. 

As a consequence of the logarithmic term in the Hadamard function
(\ref{eq:G(1)fR}), it will appear a similar term in
the stress-energy tensor. The argument of this 
logarithm depends on the curvature scalar
and thus it will grow unbounded as the flat region is approached.
However, the coefficient that will appear in front  of such a logarithm,
depends only on locally constructed curvature terms (as can be seen
from (\ref{eq:G(1)fR})). Therefore,
with an adequate matching of the space-time
geometry, this coefficient will also
smoothly vanish towards the flat space region, below $t=t_0$.
The details of such a matching, however, will not affect
the main features of the stress-energy tensor,
particularly when the singularity is approached.
 
Recall that the components of $\langle T_{\mu\nu}\rangle$ are expressed
in the orthonormal thetrad basis $(\theta _1,\theta _2,\theta _3,\theta _4)$
defined upon coordinates $(t,z,x,y)$, with $t$ and $z$ defined
in (\ref{eq:deftz}). These coordinates allow us to describe symmetrically the 
head on plane wave collision.
Thus, written in these coordinates,  the stress energy tensor is diagonal.
However, an observer at rest at the collision
center is better described by means of the {\em comoving coordinates}
$T=v+u$, $Z=v-u$, such that $Z=0$ for the observer's worldline.
Observe that the coordinate systems $(t,z)$ and $(T,Z)$ are related
by the Lorentz transformation (\ref{eq:gammabeta}). Therefore,
the stress-energy tensor in the observer's orthonormal comoving frame
will be determined by expressing $\langle T_{\mu\nu}\rangle$
in the orthonormal thetrad basis 
$\theta _{\hat 1}={g^{1/2}_{{T}{T}}}\,d{T}$,
$\theta _{\hat 2}={g^{1/2}_{{Z}{Z}}}\, d{Z}$, $\theta _3={g^{1/2}_{xx}}\, dx$, 
$\theta _4={g^{1/2}_{yy}}\, dy$, defined upon coordinates $(T,Z,x,y)$.
This can be directly achieved by transforming $\langle T_{\mu\nu}\rangle$,
expressed in the basis $(\theta _1,\theta _2,\theta _3,\theta _4)$, 
by the Lorentz transformation (\ref{eq:gammabeta}), i.e.,

\bba
\langle T_{{\hat 1}{\hat 1}}\rangle &=&
\gamma ^2\left(\langle T_{11}\rangle +\beta ^2\langle T_{22}\rangle\right),
\;\;\;\;
\langle T_{{\hat 2}{\hat 2}}\rangle =
\gamma ^2\left(\langle T_{22}\rangle +\beta ^2\langle T_{11}\rangle\right),
\nonumber\\
\langle T_{{\hat 1}{\hat 2}}\rangle &=&
\gamma ^2\beta\left(\langle T_{11}\rangle +\langle T_{22}\rangle\right).
\label{eq:TMNcomov}\eea
Observe that in comoving coordinates $(T,Z)$
the collision is not symmetrical anymore and the stress-energy tensor
acquires a non-diagonal term $\langle T_{{\hat 1}{\hat 2}}\rangle$, 
which indicates the flux of momentum in the $Z$ direction.

\section{Conclusions}
 
We have given a procedure to calculate the expectation 
value of the stress-energy tensor
of a massless scalar field in a family of  space-times representing the head on
collision of two gravitational plane waves throughout the causal
past of the collision center and in the field state which
corresponds to the physical vacuum state before the collision. 
We have considered this particular region
essentially because, as we pointed out in a previous work
\cite{dor97,dor98a}, we could introduce a suitable approximation to the
space-time metric (see Fig. 3) which not only allow us to
dramatically simplify the calculations but also to keep unchanged the
main physical features, in particular the behaviour of the
stress-energy tensor near the singularity of the
interaction region.

Unfortunately, a general expression for the stress-energy tensor has not
been found. The reason is because the approximation (\ref{eq:dshatIb})
that we have used in the line element, throughout the causal past of
the collision center, depends on which particular solution, from the
Szekeres Class of solutions, we are considering. Nevertheless, one could
expect that near the singularity the main physical effects would be 
related to the exponents of the singular $\sigma ({t})$ terms in the
line element (\ref{eq:dshatIb}). Thus, the main contribution to
the stress-energy tensor would be essentially related to these exponents and
only trivially sensitive to the remaining smooth functions.

\vskip 1.25 truecm
 
{\Large{\bf Acknowledgements}}
 
\vskip 0.5 truecm
 
\noindent
I am  grateful to R. M. Wald, R. Geroch, E. Verdaguer, A. Campos,
E. Calzetta, A. Feinstein,  
J. Iba{\~n}ez and A. Van Tonder for helpful
discussions. I am also grateful to the Physics Department of Brown
University for their hospitality and to the Grup de F\'{\i}sica
Te\`orica (IFAE) de l'Universitat Aut\`onoma de Barcelona.
This work has been partially supported by 
NSF grant PHY 95-14726 to The University of Chicago and by the Direcci\'o
General de Recerca de la
Generalitat de Catalunya through the grant 1995BEAI300165.

\appendix

\section{Some midpoint expansions}

The coefficients for the {\em midpoint
expansion} of the locally constructed Hadamard function (\ref{eq:Hada1})
are:

\bban
a_1^{(0)}&=&-R\,\left(\xi-{1\over 6}\right),\;\;\;\;
{\Delta ^{(2)}}_{\mu\nu}={1\over 12}R_{\mu\nu},
\\
& &
\\
a_2^{(0)}&=&
{1\over 2}\left({1\over 6}-\xi\right)^2\, R^2
+{1\over 6}\left({1\over 5}-\xi\right)\, {R_{;\alpha}}^\alpha
-{1\over 180}\, R^{\alpha\beta} R_{\alpha\beta}
+{1\over180}\,R^{\alpha\beta\gamma\delta}R_{\alpha\beta\gamma\delta},
\\
& &
\\
{a_1^{(2)}}_{\mu\nu}&=&
{1\over 24}\left({1\over 10}-\xi\right)\, R_{;\mu\nu}
+{1\over 120}\, {R_{\mu\nu ;\alpha}}^\alpha
-{1\over 90}\, {R^\alpha}_\mu R_{\alpha\nu}+
\\
& &
\\
& & {1\over 180}\, R^{\alpha\beta}R_{\alpha\mu\beta\nu}
+{1\over 180}\, {R^{\alpha\beta\gamma}}_{\mu}\,  
R_{\alpha\beta\gamma\nu},
\\
& &
\\
{\Delta ^{(4)}}_{\mu\nu\rho\tau}&=&
 {3\over 160}\, R_{\mu\nu ;\rho\tau}
+{1\over 288}\, R_{\mu\nu}R_{\rho\tau}
+{1\over 360}\, {{{R^\alpha}_\mu}^\beta}_\nu  
R_{\alpha\rho\beta\tau}.
\eean

\vskip 1.25 truecm
 
{\Large{\bf Figure captions}}
 
\vskip 0.5 truecm
 
\noindent
{\bf Fig. 1}
The colliding plane wave space-time consists of two
approaching waves, regions II and III, in a flat background, region
IV, and an interaction region, region I. The two waves move in the
direction of two null coordinates $u$ and $v$.
The four space-time regions
are separated by the two null wave fronts $u=0$ and $v=0$. The
boundary between regions I and II is  $\{0\leq u<1,\; v=0\}$, the
boundary between regions I and III is $\{u=0,\; 0\leq v <1\}$, and
the boundary of regions II and III with region IV is
$\Sigma =\{u\leq 0,\;v=0\}\cup\{u=0,\;v\leq 0\}$. Region I meets region IV
only at the surface $u=v=0$. The singularity in the region
I corresponds to the hypersurface $u^{n_1}+v^{n_2}=1$ and plane wave regions
II and III meet such a singularity
only at ${\cal P}=\{u=1,\; v=0\}$ and
${\cal P}'=\{u=0,\; v=1\}$ respectively. 
The hypersurfaces  $u=1$ in region II  and $v=1$ 
in region III are a type of topological singularities commonly referred as
folding singularities and they
must be identified with $\cal P$ and ${\cal P}'$ respectively.

\vskip 0.5 truecm

\noindent
{\bf Fig. 2}
The subset of Cauchy data which affects the evolution of the 
quantum field along the center $u=v$ of the plane wave  collision lies on
the segments 
${\hat\Sigma}_{\rm I}=\{0\leq u<{\hat s},\; v=0\}\cup
\{u=0,\; 0\leq v <{\hat s}\}$,
where ${\hat s}^{n_1}+{\hat s}^{n_2}=1$.
Region $\cal S$ is the causal future of this Cauchy data (or
equivalently, the causal past of the collision center).

\vskip 0.5 truecm

\noindent
{\bf Fig. 3}
We change the {\em mode propagation problem} for the plane wave
collision, in the causal past of the collision center, region $\cal S$,
by a much simpler Schr\"odinger-type problem which consists in: 
i) performing a Lorentz transformation in order to work with an adequate
set of coordinates $(t,z)$,
ii) eliminating the dependence of the field equation on coordinate
$z$ by taking $z=0$
and
iii) substituting the Cauchy data, which
come from the single plane wave regions, on segments
${\hat\Sigma}_{\rm I}=\{0\leq u<{\hat s},\; v=0\}\cup
\{u=0,\; 0\leq v <{\hat s}\}$ 
by much simpler Minkowski Cauchy data. This procedure is essentially
equivalent to:
i) writing the line element in coordinates $(t,z)$,
ii) modifying the space-time geometry in the causal
past of the collision center by eliminating
the dependence on coordinate $z$ in the
line element, setting $z=0$ 
and 
iii) smoothly matching this
line element, through plane wave regions II and III, 
with the flat spacetime below the segment
$\{{T}=0,\; -{\hat s}< Z < {\hat s}\}$, where we denote $T=u+v$, $Z=v-u$.

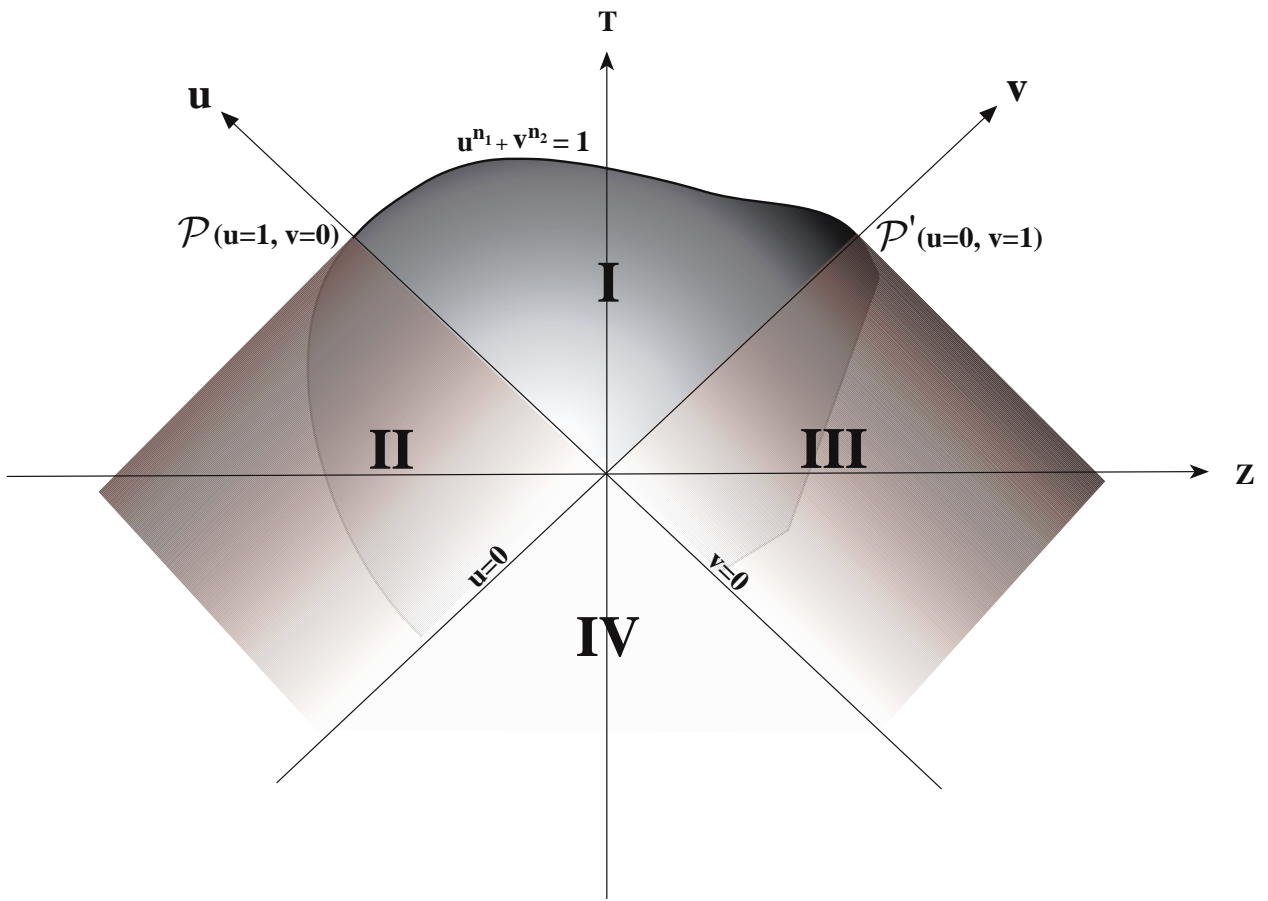
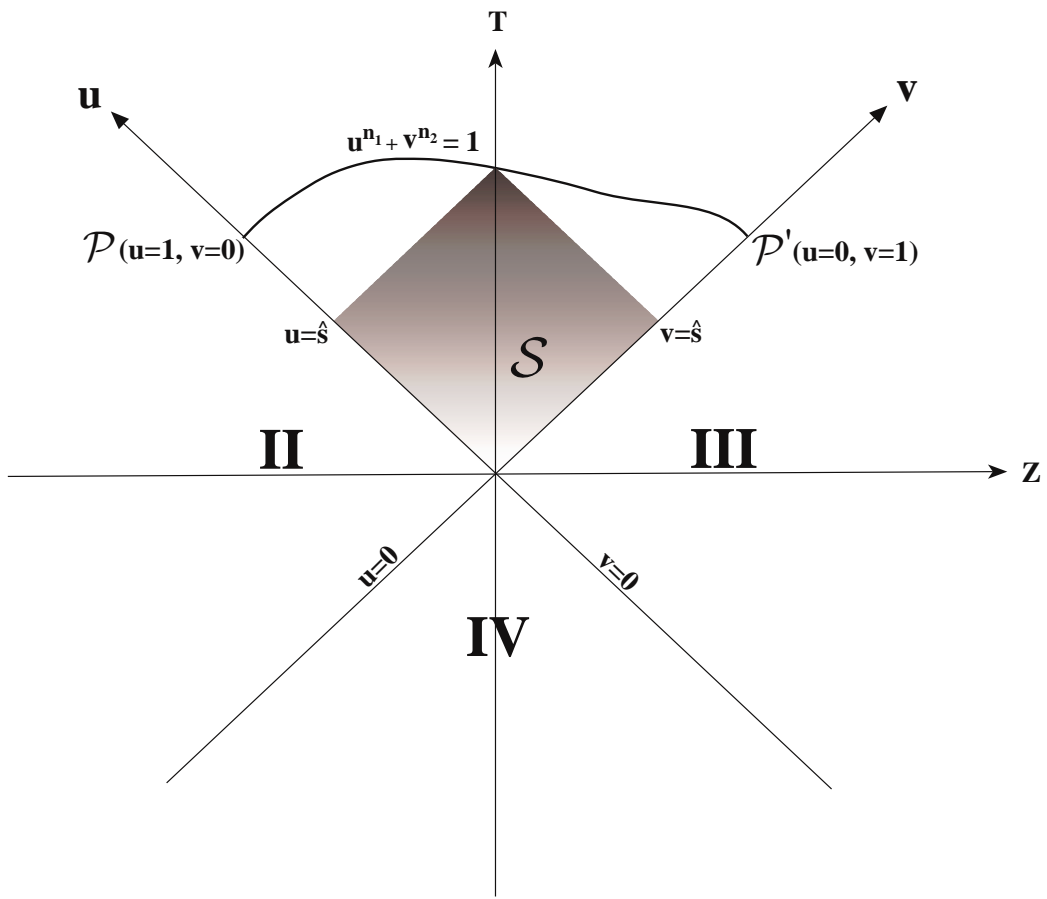
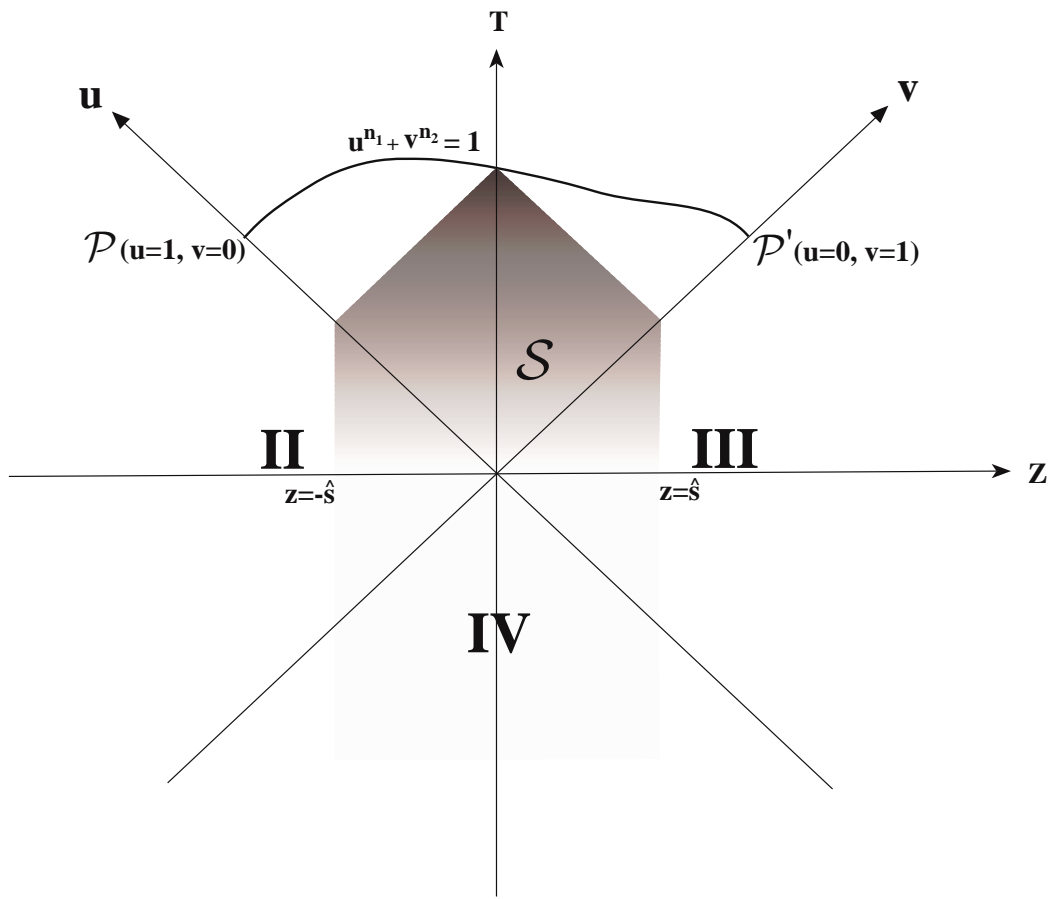
\end{document}